\documentstyle[preprint,eqsecnum,aps,epsfig]{revtex}
\preprint{SNUTP 96-005}

\begin{document}
\draft
\title{
Method for Detecting Berry's Phase
in a Modified Paul Trap
}

\author{Jong-Chan Park,
Jeong-Young Ji,\footnote{
Electronic address: jyji@phyb.snu.ac.kr}
and
Kwang-Sup Soh\footnote{
Electronic address: kssoh@phyb.snu.ac.kr}
}
\address{Department of Physics Education,
Seoul National University, Seoul 151-742, Korea}
\maketitle

\begin{abstract}
~We modify the time-dependent electric potential of the Paul trap
from a sinusoidal waveform to a square waveform.
The exact quantum motion and the Berry's phase of an electron
in the modified Paul trap are found
in an analytically closed form.
We consider a scheme to detect the Berry's phase
by a Bohm-Aharonov-type interference experiment
and point out a critical
property which renders it practicable.

\end{abstract}
\pacs{03.65.Bz}

\section{Introduction}

The Paul trap is an instrument which suspends free charged and neutral
particles without material walls. Such traps permit the observation of
isolated particles, even a single one, over a long period of
time~\cite{Paul90}.
The Hamiltonian of the Paul trap  has the form of
a time-dependent harmonic oscillator,
\begin{equation}
H (t) = \frac{1}{2 m} p^{2} (t) + \frac{1}{2} m \omega^{2} (t) q^{2} (t)
\label{H(t)}
\end{equation}
whose effective spring constant is of the form~\cite{Brown91}
\begin{equation}
k (t) = a + b \cos (2 \pi t/\tau).
\label{eff:k}
\end{equation}
The quantum motion of the Paul trap has been studied
in Refs.~\cite{Brown91,FengW95,Ji95:Heis}.
It is well known that the generalized invariant for Eq.~(\ref{H(t)})
can be written as~\cite{LR69}
\begin{equation}
I(t) = g_{-}(t) \frac{p^{2}}{2} +
g_{0} (t) \frac{{ p q + q p }}{2} +
g_{+} (t) \frac{q^{2}}{2} .
\end{equation}
Here, using the classical solutions
satisfying
\begin{equation}
{\ddot{f}}_{1,2} (t) + \omega^{2} (t) f_{1,2} (t) = 0,
\label{EOM}
\end{equation}
we have~\cite{Ji95:Heis}
\begin{eqnarray}
g_{-} (t) &=&
c_{1} f_{1}^{2} (t) +
c_{2} f_{1} (t) f_{2} (t) +
c_{3} f_{2}^{2} (t) ,
\nonumber \\
g_{0~} (t) &=& - m
\{
c_{1} f_{1} (t) {\dot{f}}_1 (t)
\nonumber \\
& & +
(c_{2} / 2) [ {\dot{f}}_1 (t) f_{2} (t) +
f_{1} (t) {\dot{f}}_2 (t) ]
\label{gi:ci} \\
&& + c_{3} f_{2} (t) {\dot{f}}_2 (t) \},
\nonumber \\
g_{+} (t) &=&
m^2 [
c_{1} \dot{f}_{1}^{2} (t) +
c_{2} \dot{f}_{1} (t) \dot{f}_{2} (t)+
c_{3} \dot{f}_{2}^{2} (t) ]
\nonumber
\end{eqnarray}
where $c_{1}$, $c_{2}$, and $c_{3}$ are arbitrary constants.

Recently, Ji et al.~\cite{Ji95:Exac} found
the exact eigenfunctions of $I(t)$:
\begin{eqnarray}
\psi_{n} (q,t) &=& \frac{1}{ \sqrt{2^{n} n! }}
\left(
\frac{\omega_{I}}{\pi g_{-} (t) }
\right)^{\frac{1}{4}}
e^{ -i \frac{{g_{0} (t) }}{ 2 g_{-} (t) } q^{2} } \nonumber  \\
& &\times  e^{ -i \int dt \frac{\omega_{I}}{m g_{-} (t) } \left(n
+ \frac{1}{2} \right) }
e^{- \frac{\omega_{I}}{2 g_{-} (t)}  q^{2} }
H_{n} \left( \sqrt{ \frac{{\omega_{I} }}{g_{-} (t) } } q \right)
\label{WF}
\end{eqnarray}
where $H_n$ is Hermite polynomial.
For a time-periodic quantum harmonic oscillator,
analyzing the wave function in Eq.~(\ref{WF}),
they constructed a cyclic initial state (CIS) such that
$\psi_{n} (t + \tau') = e^{i \chi_{n} (\tau') } \psi_{n} (t)$
with
\begin{equation}
\chi_{n} (\tau' ) = -\left(n + \frac{1}{2} \right) \int_{0}^{\tau'}
\frac{\omega_{I}}{m g_{-} (t) } dt .
\label{BP}
\end{equation}
and calculated the corresponding Berry's phase
(see Ref.~\cite{Berry84} for the Berry's phase and
Ref.~\cite{AharonovA87} for its nonadiabatic generalization).
Subsequently, a new type of CIS,
whose period is a multiple of the period of the Hamiltonian,
was found ~\cite{Ji95:Ntau}.

In this paper,
we modify the time-periodic electric potential
from the sinusoidal waveform in Eq.~(\ref{eff:k})
to a square waveform. This square potential has
stable classical solutions, as the sinusoidal potential does.
This means that
we can suspend charged particles
using this modified potential,
as we do in the original Paul trap.
Furthermore, the classical solutions of this modified Paul trap are
very simple, so we can calculate the exact
quantum solutions in a simple closed form.
(Note that the classical solutions of the original Paul trap
are Mathieu functions, which are difficult to deal with.)
The purpose of this paper is to find the
Berry's phase for the modified Paul trap and
to propose an experimental scheme to detect it.

As seen from Eq.~(\ref{gi:ci}),
there is an arbitrariness in fixing the invariant, and
hence the complete set of the Fock space
(eigenstates of the invariant).
Therefore, we should show that
the phase change of an eigenstate, Eq.~(\ref{WF}),
is irrelevant to which invariant we choose.
There is another problem: When we let the
electron beam pass through the modified Paul trap,
it seems that we should have a single eigenstate for
a coherent interference pattern.
However, it turns out that
if we prepare a plane wave of the electron
-- which can be expanded as the eigenstates in Eq.~(\ref{WF}) --
we get a coherent interference pattern.

In Sec.~\ref{EQSPT},
we apply the result of Refs.~\cite{Ji95:Heis}, \cite{Ji95:Exac}, and \cite{Ji95:Ntau}
to the modified Paul trap to find the exact quantum state
and the Berry's phase.
In Sec.~\ref{EDBP},
we present a Bohm-Aharonov-type experimental method
for detecting the Berry's phase of this system.
The key feature which renders this experiment practicable is
that
the phase change is independent of the invariant we choose,
and for a coherent interference pattern
it is sufficient to prepare a plane wave entering the trap.
A summary and discussions are given in the last section.

\section{Exact quantum motion of the modified Paul trap}
\label{EQSPT}

\subsection{Quantum Mechanics of the Paul Trap}

The classical and quantum motion of an electron in the Paul trap is
described by the following Hamiltonian~\cite{Paul90}:
\begin{equation}
H(t) = H_{x}(t) + H_{y}(t) + H_{z}(t)
\end{equation}
where
\begin{mathletters}
\label{Hi}
\begin{eqnarray}
H_{x} &=& \frac{1}{2m} p_{x}^{2}
+ \frac{1}{2} m \omega_x^2 x^{2} ,
\label{H:x} \\
H_{y} &=& \frac{1}{2m} p_{y}^{2}
+ \frac{1}{2} m \omega_y^2  y^{2},
\label{H:y} \\
H_{z} &=& \frac{1}{2m} p_{z}^{2} .
\label{H:z}
\end{eqnarray}
\end{mathletters}

\noindent
Here, the Hamiltonians of the $x$- and the $y$-motions
have the form of a time-dependent harmonic oscillator
with
\begin{equation}
\omega_x^2 = \frac{e \Phi(t)}{m d^{2}} = -\omega_y^2
\label{freq}
\end{equation}
where
\begin{equation}
\Phi (t)= U + V \cos (2 \pi t / \tau).
\label{ap-Vol}
\end{equation}
is an applied voltage,
$d$ is the gap of the walls of the Paul trap,
and $e$ is the absolute value of the electron charge.

The wave function of this system satisfies the time-dependent
Schr\"odinger equation
\begin{equation}
i \frac{{\partial}}{ \partial t} \Psi (x,y,z,t)
= H(t) \Psi (x,y,z,t) .
\label{total-Sch}
\end{equation}
Using the method of separation of variables,
we have three independent equations:
\begin{equation}
i \frac{{\partial}}{ \partial t} \Psi_{i} ({\rm r}_i, t)
= H_{i} (t) \Psi_{i} ({\rm r}_i,t),~
(i = x,y,z).
\label{Sch-i}
\end{equation}
Here, the equation in the $z$-direction gives the plane-wave solution
$\Psi_{z} (z,t) = e^{i(k_z z - E_{z} t)}$.
In addition, since Eqs.~(\ref{H:x}) and (\ref{H:y})
are the Hamiltonian of a time-dependent
harmonic oscillator,
we can find $\Psi_{x}$ and $\Psi_{y}$
using the methods found in Refs.~\cite{Ji95:Heis} and \cite{Ji95:Exac}.

\subsection{Modified Paul Trap}

\begin{figure}[htb]
\centerline{\epsfig{figure=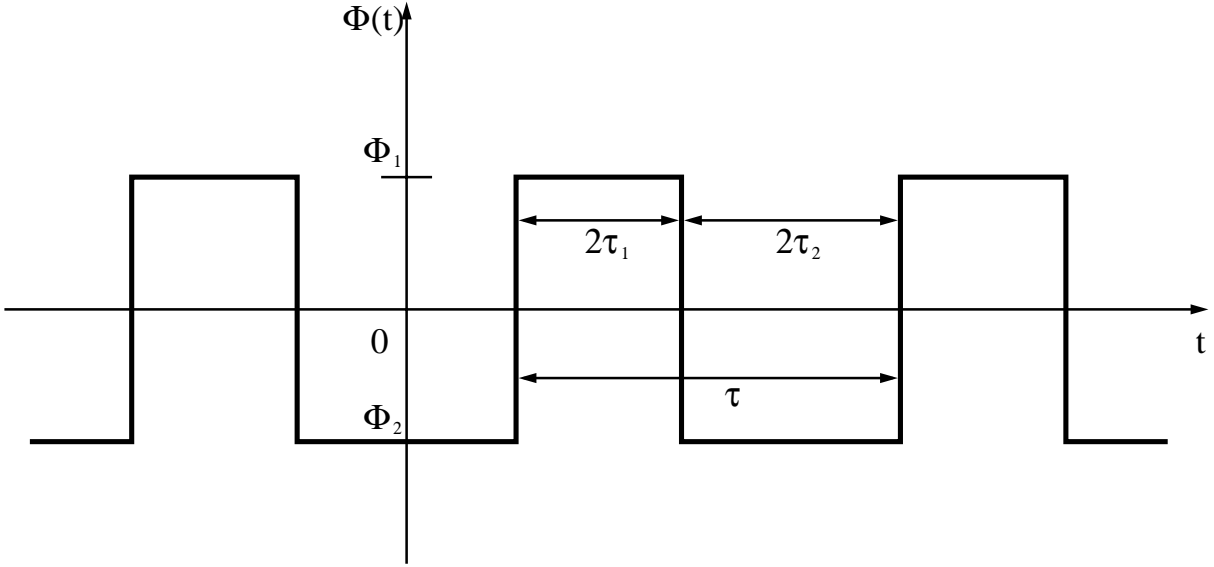,width=\hsize,angle=0}}
\caption{The time-dependent potential of a square waveform.
}
\label{fig1}
\end{figure}

Now, we modify the applied voltage from the form in Eq.~(\ref{ap-Vol}) to the
following square wave form (see Fig.~\ref{fig1}):
\begin{equation}
\Phi(t) =
{
\left\{\begin{array}{ll}
\Phi_1 > 0 , &
{\rm for}~\tau_{2} - \tau < t - r \tau < - \tau_{2}, \\
\Phi_2 < 0 , &
{\rm for}~- \tau_{2} < t - r \tau < \tau_{2},
\end{array}\right.
}\label{sq-Pot}
\end{equation}
where $r$ is an integer.
Then, the frequencies of $H_x$ and $H_y$ are described by
\begin{mathletters}
\label{omx-omy}
\begin{eqnarray}
\omega_x^2 (t) &=&
{
\left\{\begin{array}{ll}
\omega_{1}^2 , &
{\rm for}~ \tau_{2} - \tau < t - r \tau < - \tau_{2}, \\
- \omega_{2}^2 , &
{\rm for}~ - \tau_{2} < t - r \tau < \tau_{2} ,
\end{array}\right.}
\label{om:x}\\
\omega_y^2 (t) &=&
{
\left\{\begin{array}{ll}
- \omega_{1}^2 , &
{\rm for}~ \tau_{2} - \tau < t - r \tau < - \tau_{2}, \\
\omega_{2}^2 , &
{\rm for}~ - \tau_{2} < t - r \tau < \tau_{2} ,
\end{array}\right.}
\label{om:y}
\end{eqnarray}
\end{mathletters}
where
\begin{equation}
\omega_1^2 = \frac{e |\Phi_1| }{2 m d^2},~
\omega_2^2 = \frac{e |\Phi_2| }{2 m d^2}.
\label{freq12}
\end{equation}

In order to study the quantum mechanics of this system,
it is necessary to determine independent classical solutions of
Eq.~(\ref{EOM}) for the $x$- and the $y$-components
with Eqs.~(\ref{om:x}) and (\ref{om:y}), respectively.
These solutions are fully analyzed in Ref.~\cite{Ji95:Ntau}.
Since the effective spring constant of the modified Paul trap
alternates between positive and {\it negative} values,
we should check carefully that
the solutions of Ref.~\cite{Ji95:Ntau}
are applicable in this model.
After tedious calculations, we verified that
our classical solutions are
identical with the solutions of Ref.~\cite{Ji95:Ntau}
with the replacements of
$\omega_2$ by $-i \omega_2$ in the $x$-component and
$\omega_1$ by $-i \omega_1$ in the $y$-component.

As a result, we find the classical solutions for the $x$-component to be
\begin{equation}
f_x (t) =
{ \left\{\begin{array}{ll}
 A_{x,r} e^{i \omega_{1} (t - r \tau)}+
 B_{x,r} e^{-i \omega_{1} (t - r \tau)} ,
  & {\rm for}~ \tau_{2} - \tau < t - r \tau < - \tau_{2}, \\
  C_{x,r} e^{ \omega_{2} (t - r \tau)}+
  D_{x,r} e^{- \omega_{2} (t - r \tau)} ,
  & {\rm for}~ - \tau_{2} < t - r \tau < \tau_{2},
\end{array}\right.
} \label{genSol}
\end{equation}
where
\begin{equation}
{ \left(\begin{array} {c}
C_{x,r} \\ D_{x,r} \end{array}\right)} =
\frac{1}{2} { \left(\begin{array} {cc}
(1+i\omega_{1}/\omega_{2})e^{-i\omega_{1}\tau_{2}+\omega_{2}\tau_{2}}&
(1-i\omega_{1}/\omega_{2})e^{ i\omega_{1}\tau_{2}+\omega_{2}\tau_{2}}\\
(1-i\omega_{1}/\omega_{2})e^{-i\omega_{1}\tau_{2}-\omega_{2}\tau_{2}}&
(1+i\omega_{1}/\omega_{2})e^{ i\omega_{1}\tau_{2}-\omega_{2}\tau_{2}}
  \end{array}\right)}
{ \left(\begin{array} {c}
A_{x,r} \\ B_{x,r} \end{array}\right)} .
\end{equation}
The coefficients $A_{x,r}$ and $B_{x,r}$,
belonging to successive values of
$r$, can be related by a matrix $P$
obtained by imposing continuity for
$f_x (t)$ and its derivative at $t=-\tau_{2}+r\tau$ and
$t=\tau_{2}+r\tau$.
These lead to
\begin{equation}
{ \left(\begin{array} {c}A_{x,r} \\ B_{x,r} \end{array}\right)} =
P^r { \left(\begin{array} {c}
A_{x,0} \\ B_{x,0} \end{array}\right)} ,
\end{equation}
with
\begin{equation}
P={\left(\begin{array}{cc}
				(\alpha_{x,1}-i \beta_{x,1})e^{i\omega_{1}\tau} &
                -i\beta_{x,2}e^{i\omega_{1}\tau} \\
                i \beta_{x,2}e^{-i\omega_{1}\tau} &
                (\alpha_{x,1}+i \beta_{x,1})e^{-i \omega_{1}\tau}
   \end{array}\right)},
\end{equation}
where
\begin{mathletters}
\begin{eqnarray}
\alpha_{x,1} &=& \cos 2 \omega_{1} \tau_{2} \cosh 2 \omega_{2} \tau_{2} +
\frac{\eta}{2} \sin 2 \omega_{1} \tau_{2} \sinh 2 \omega_{2} \tau_{2} , \\
\beta_{x,1} &=& \sin 2 \omega_{1} \tau_{2} \cosh 2 \omega_{2} \tau_{2} -
\frac{\eta}{2} \cos 2 \omega_{1} \tau_{2} \sinh 2 \omega_{2} \tau_{2} , \\
\beta_{x,2} &=& \frac{\epsilon}{2} \sinh 2 \omega_{2} \tau_{2} ,
\end{eqnarray}
\end{mathletters}
and
\begin{equation}
\epsilon = \frac{\omega_{1}}{\omega_{2}} + \frac{\omega_{2}}{\omega_{1}} ,~
\eta = \frac{\omega_{1}}{\omega_{2}} - \frac{\omega_{2}}{\omega_{1}} ,
\end{equation}
where $\alpha_{x,1}$, $\beta_{x,1}$, and $\beta_{x,2}$ satisfy the condition
\begin{equation}
\alpha_{x,1}^{2} + \beta_{x,1}^{2} - \beta_{x,2}^{2} =1 .
\label{albe}
\end{equation}

Solving the eigenvalue problem for the matrix $P$, we find the eigenvalues
\begin{equation}
p_{\pm} = \lambda_x \pm \sqrt{\lambda_x^{2} - 1}
\end{equation}
where
$\lambda_x = \alpha_{x,1} \cos \omega_{1} \tau +
\beta_{x,1} \sin \omega_{1} \tau , $
and their corresponding eigenvectors
\begin{equation}
{\left(\begin{array} {c}A_{x,0} \\ B_{x,0} \end{array}\right)} \propto
{\left(\begin{array}{c}
        \beta_{x,2} e^{i \omega_{1} \tau} \\
		\nu_x\pm i\sqrt{\lambda_x^{2} - 1}\end{array}\right)}
\end{equation}
where
$\nu_x=\alpha_{x,1}\sin\omega_{1}\tau-\beta_{x,1}\cos\omega_{1}\tau$.
If $|\lambda_x|\leq 1$, $p_{\pm} $ are complex conjugates. Investigating the
form of the matrix $P$, it is easy to find that the solutions corresponding
to two eigenvalues are also complex conjugates. Therefore, the two independent
solutions are taken to be the real and the imaginary parts of one of them.
In this case, we can set
\begin{equation}
A_{x,0} = \beta_{x,2} e^{i \omega_{1} \tau} ,~
B_{x,0} = \nu_x - \sqrt{1-\lambda_x^{2}}
\end{equation}
with one eigenvalue
\begin{equation}
p_{+} = \lambda_x + i \sqrt{1-\lambda_x^{2}} = e^{i \phi_x}
\end{equation}
where $\tan \phi_x = \sqrt{1 - \lambda_x^{2}} / \lambda_x . $
Then, the classical solution for $|\lambda_x| \leq 1 $ can be written as
\begin{equation}
f_x (t) =
{ \left\{\begin{array}{ll}
e^{i r \phi_x}\left[
A_{x,0} e^{i \omega_{1} (t - r \tau)} +
B_{x,0} e^{-i \omega_{1} (t - r \tau)}\right] ,&
{\rm for}~ \tau_{2} - \tau < t - r \tau < - \tau_{2}, \\
e^{i r \phi_x}\left[
C_{x,0} e^{ \omega_{2} (t - r \tau)} +
D_{x,0} e^{- \omega_{2} (t - r \tau)} \right] ,&
{\rm for}~ - \tau_{2} < t - r \tau < \tau_{2}.
\end{array}\right.}
\label{stSol:x}
\end{equation}
In the same way, we have the classical solution
for the $y$-component:
\begin{equation}
f_y (t) =
{\left\{\begin{array}{ll}
e^{i r \phi_y}\left[
A_{y,0} e^{ \omega_{1} (t - r \tau)} +
B_{y,0} e^{- \omega_{1} (t - r \tau)}\right] ,&
{\rm for}~\tau_{2} - \tau < t - r \tau < - \tau_{2}, \\
e^{i r \phi_y}\left[
C_{y,0} e^{i \omega_{2} (t - r \tau)} +
D_{y,0} e^{-i \omega_{2} (t - r \tau)} \right] ,&
{\rm for}~ - \tau_{2} < t - r \tau < \tau_{2},
\end{array}\right.}
\label{stSol:y}
\end{equation}
with
\begin{equation}
A_{y,0}=\beta_{y,2}e^{ \omega_{1} \tau},~
B_{y,0}=\nu_y-i\sqrt{1-\lambda_y^{2}},
\end{equation}
\begin{equation}
{\left(\begin{array} {c}
C_{y,0} \\ D_{y,0} \end{array}\right)} =
\frac{1}{2} { \left(\begin{array} {cc}
(1-i\omega_{1}/\omega_{2})e^{-\omega_{1}\tau_{2}+i\omega_{2}\tau_{2}}&
(1+i\omega_{1}/\omega_{2})e^{ \omega_{1}\tau_{2}+i\omega_{2}\tau_{2}}\\
(1+i\omega_{1}/\omega_{2})e^{-\omega_{1}\tau_{2}-i\omega_{2}\tau_{2}}&
(1-i\omega_{1}/\omega_{2})e^{ \omega_{1}\tau_{2}-i\omega_{2}\tau_{2}}
\end{array}\right)}
{\left(\begin{array}{c}
A_{y,0} \\ B_{y,0}
\end{array}\right)},
\end{equation}
where
\begin{equation}
\nu_y=\alpha_{y,1}\sinh\omega_{1}\tau-\beta_{y,1}\cosh\omega_{1}\tau ,
\end{equation}
\begin{mathletters}
\begin{eqnarray}
\alpha_{y,1} &=& \cosh 2 \omega_{1} \tau_{2} \cos 2 \omega_{2} \tau_{2} -
\frac{\eta}{2} \sinh 2 \omega_{1} \tau_{2} \sin 2 \omega_{2} \tau_{2} , \\
\beta_{y,1} &=& \sinh 2 \omega_{1} \tau_{2} \cos 2 \omega_{2} \tau_{2} -
\frac{\eta}{2} \cosh 2 \omega_{1} \tau_{2} \sin 2 \omega_{2} \tau_{2} , \\
\beta_{y,2} &=& \frac{\epsilon}{2} \sin 2 \omega_{2} \tau_{2} ,
\end{eqnarray}
\end{mathletters}
and
\begin{equation}
 e^{i \phi_y} = \lambda_y+i\sqrt{1-\lambda_y^{2}}
\end{equation}
where
$\lambda_y=\alpha_{y,1}\cosh\omega_{1}\tau-\beta_{y,1}\sinh\omega_{1}\tau.$

The two independent real solutions $f_{1}(t)$ and $f_{2}(t)$ are
given by
\begin{equation}
f_{1} (t) = \frac{1}{2}  [f(t) + f^{*} (t)],~
f_{2} (t) = \frac{1}{2i} [f(t) - f^{*} (t)]
\end{equation}
for the $x$ and the $y$ components, respectively.
These solutions exhibit stable motions for $\lambda\leq 1$
($\lambda$ stands for $\lambda_x$ or $\lambda_y$);
that is, they oscillate with bounded amplitudes.

\begin{figure}[htb]
\centerline{\epsfig{figure=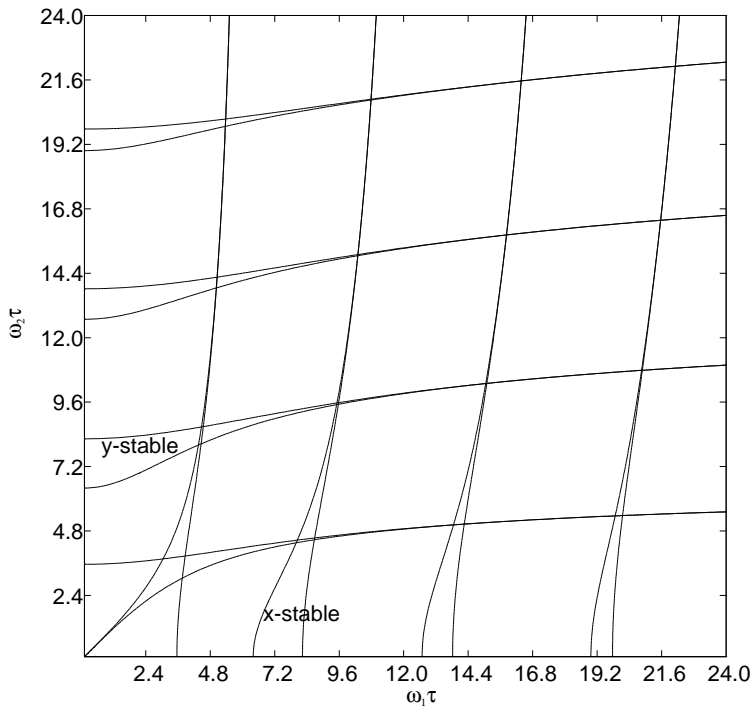,width=100mm,angle=0}}
\caption{The stability-instability diagram.
The vertical strips stand for the stable regions in the $x$-motion,
the horizontal strips for the $y$-motion.
}
\label{fig2}
\end{figure}

It is important to know the stable regions
in $\omega_1$-$\omega_2$ diagram
where the classical solutions are stable.
For $\tau_1 = \tau_2$, we present the stable regions and the unstable regions
in Fig.~\ref{fig2}.
This map is similar to the stability diagram obtained
from the Mathieu equation. Only the overlapping regions
of the $x$-stable and the $y$-stable regions
are of our interest. Therein, the motion
is stable both in the $x$-direction and the $y$-direction.
On the other hand, when $|\lambda|>1$, the solutions diverge at
$t\rightarrow\infty$ or $t\rightarrow-\infty$
as discussed in Ref.~\cite{Ji95:Ntau}.

Now, we fix the generalized invariant by
fixing $c_1$, $c_2$, and $c_3$ in Eq.~(\ref{gi:ci}).
For example, to find $I(t)$ such that $I(t) = H(t_0)$
in the $x$-component ($t_0$ denotes the initial time),
we fix those three parameters as~\cite{Ji96:Temp}
\begin{eqnarray}
c_{x,1} &=&
\frac
{{\beta}_{x,2}^2+B_{x,0}^2-2{\beta}_{x,2}B_{x,0}\cos\omega_1\tau}
{m{\left({\beta}_{x,2}^2-B_{x,0}^2\right)}^2},
\nonumber \\
c_{x,2} &=&
\frac
{-4{\beta}_{x,2}B_{x,0}\sin\omega_1\tau}
{m{\left({\beta}_{x,2}^2-B_{x,0}^2\right)}^2} ,
\label{const} \\
c_{x,3} &=& \frac
{{\beta}_{x,2}^2+
B_{x,0}^2+2{\beta}_{x,2}B_{x,0}\cos\omega_1\tau}
{m{\left({\beta}_{x,2}^2-B_{x,0}^2\right)}^2}.
\nonumber
\end{eqnarray}
In this way, we can get the exact wave function of
the modified Paul trap, Eq.~(\ref{WF}),
and the phase change (which includes
the Berry's phase) for a period, Eq.~(\ref{BP}).

\section{Experimental method of detecting Berry's phase}
\label{EDBP}

\subsection{$N\tau$-periodic Wave Function}

In this section, we present an experimental
method to detect the effect of the Berry's phase.
The existence of the CIS is
provided by the periodic classical solutions.
As discussed in Refs.~\cite{Ji95:Exac} and \cite{Ji95:Ntau},
if it holds in Eqs.~(\ref{stSol:x}) and (\ref{stSol:y}) that
\begin{equation}
\phi = \frac{l}{N'} 2 \pi ~(l, N' = {\rm  integers~ and}~ N' \neq 0)
\label{pe-cond}
\end{equation}
(where the index of $x$ and $y$ is understood),
the classical solution is $N' \tau$-periodic.
Then, $g_- (t)$ is $N'\tau/\epsilon$-periodic
($\epsilon=1$ for odd $N'$, $\epsilon=2$ for even $N'$), accordingly,
so is the wave function in Eq.~(\ref{WF}).

When we have two independent real classical solutions,
say $f_1(t)$ and $f_2(t)$,
we can always construct the complex solution as
\begin{equation}
f_c (t)= d_1 f_1(t)+ (d_2 + i d_3) f_2(t),
\end{equation}
where $d_1$, $d_2$, and $d_3$ ($d_1 d_3 \neq 0$) are real parameters.
This solution can be written in polar form~\cite{MagnusW66}:
\begin{equation}
f_c (t)=|f_c (t)|e^{i \theta(t)}
\label{polar-f}
\end{equation}
where
\begin{equation}
\theta(t)=\int_0^{t}\frac{\omega_{I}}{m |f_c (t')|^2}dt'.
\end{equation}
If we set $c_1=d_1^2$, $c_2=2 d_1 d_2$, and $c_3=d_2^2+d_3^2$,
we have $g_-(t)=|f_c(t)|^2$.
Then, the quantum phase, Eq.~(\ref{BP}), of the $n$-th eigenstate,
which is also a CIS with a period $\tau'$, can be rewritten as
\begin{equation}
\chi_{n} (\tau')=-\left(n+\frac{1}{2}\right)\theta(\tau').
\label{qp:cp}
\end{equation}

Now we are ready to prove that the quantum phase in Eq.~(\ref{qp:cp})
is independent of the choice of the invariant.
That is, the phase change of the eigenfunction of the invariant
does not depend on what values of $c_i$ we choose.
The proof is as follows:
If we assume that the phase change of the classical solution in Eq.~(\ref{polar-f})
is altered by varying the parameter values $d_i(i=1,2,3)$ or $c_i$,
then there are many classical solutions
corresponding to the respective periods.
However, this contradicts the fact that
the classical solution of Eq.~(\ref{EOM}) has only two independent solutions
and that they are the real and the imaginary parts of Eq.~(\ref{polar-f}).
This completes our proof.

\subsection{Experimental Setting}

\begin{figure}[htb]
\centerline{\epsfig{figure=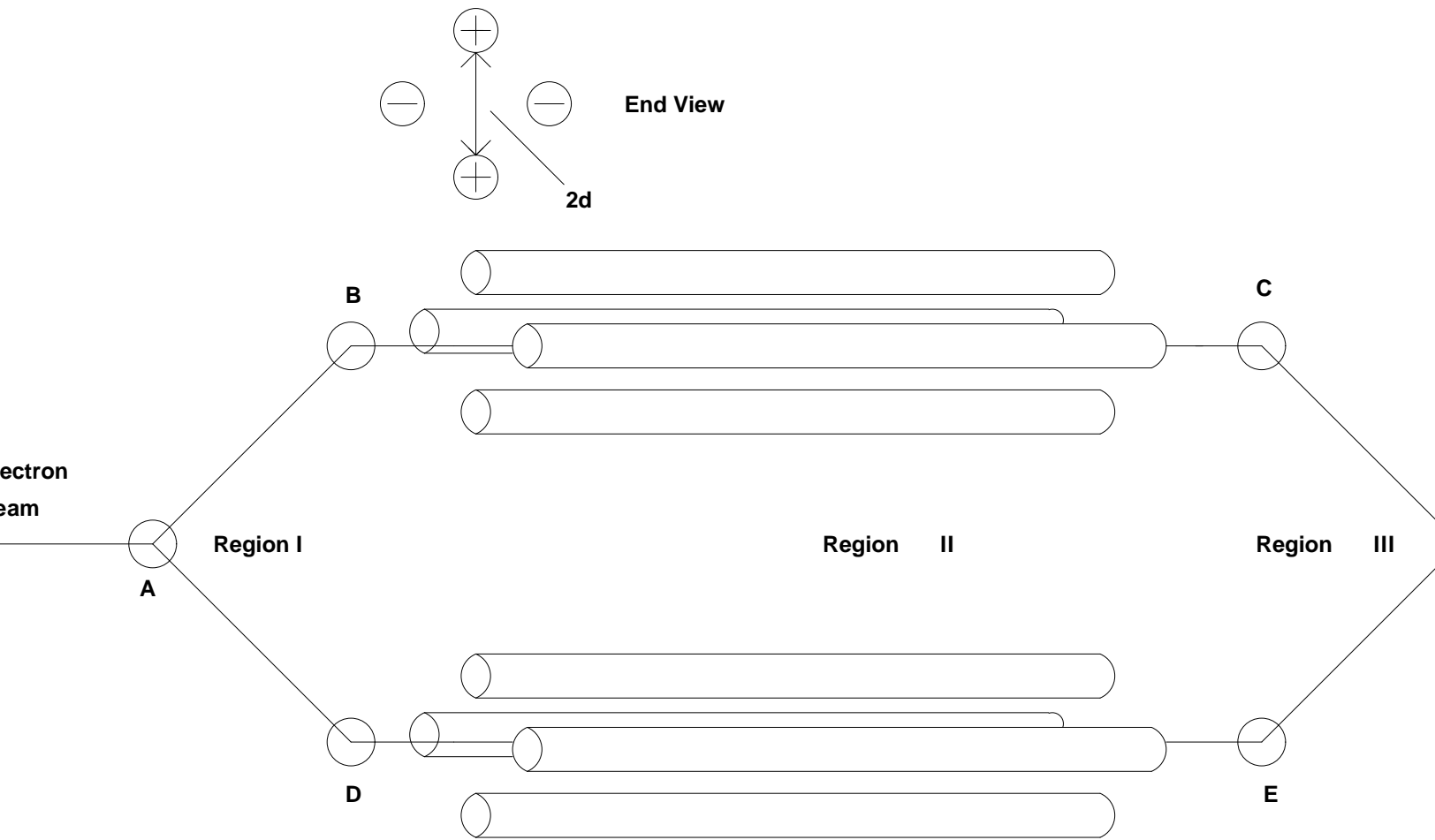,width=140mm,angle=-90}}
\caption{The schematic diagram of
the experimental arrangement.
}
\label{fig3}
\end{figure}

Now let us consider the experimental arrangement.
Suppose we have a single coherent electron beam which is split into
two parts, and suppose each part is allowed to enter the modified Paul trap,
as shown in Fig.~\ref{fig3}
This experiment is similar to the experiment
illustrated by Aharonov and Bohm in Ref.~\cite{AharonovB59}.
After the beams pass through the modified Paul traps,
they are combined to interfere coherently at the point F.
Let us denote the paths A-B-C-F and A-D-E-F
by path 1 and 2, respectively.
The electric potential vanishes in region I so that
the wave function of the electron is described by
a plane wave which propagates along the $z$-direction:
$\Psi({\rm I}) = L^{-3/2} e^{ik_z z - E_z t}$
where $L$ is a suitable normalization factor.
In region II, the potential varies as a function
of time according to Eq.~(\ref{sq-Pot}),
$\Phi(t)$ and $-\Phi(t)$ in the $x$- and the $y$-directions, respectively,
but $\Phi_1$ and/or $\Phi_2$ have different values
for paths 1 and 2.
When the electron is in region III,
the potential vanishes again.

Let $ \Psi_1 $ and $\Psi_2$ be the wave functions that
pass through path 1 and path 2, respectively.
In region I, $\Psi({\rm I})=\Psi_1({\rm I})=\Psi_2({\rm I})$.
Then, as they enter region II,
the two wave functions suffer different potentials
in the two different Paul traps.
Finally, we have
$\Psi({\rm III})=\Psi_1({\rm III})+\Psi_2({\rm III})$ in region III.

In order to form a sharp interference pattern,
it is necessary to have the wave functions at point F
be of the form
\begin{equation}
\Psi({\rm III})=
\Psi({\rm I})e^{-i\Theta_1} +
\Psi({\rm I})e^{-i\Theta_2},
\label{sharp}
\end{equation}
such that the pattern depends upon the phase difference
$\Theta_2 - \Theta_1$.
We emphasize that the critical factor which renders
our experimental scheme practical is that
any plane wave which splits at the point A does
interfere at F, as required, in the form of Eq.~(\ref{sharp}).

Firstly, when $\Psi_{1x}$ is a single eigenstate of the LR invariant,
the phase change can be easily obtained by using Eq.~(\ref{BP}), and
it is evident the final wave function at F
is of the form of Eq.~(\ref{sharp}). Next,
for a plane wave propagating along the $z$-direction with
the wave number ${\bf k}$,
\begin{equation}
{\bf v}=\frac{\hbar {\bf k}}{m}=\frac{D}{T} \hat{z}
\label{w-n}
\end{equation}
gives a final wave function of the form of Eq.~(\ref{sharp}),
as we will see below.
Here, $D$ is the length of the Paul trap, and
$T$ is a multiple of the minimal period of the CIS,
which can be controlled by the applied voltage.
In this situation, the wave function of the electron
entering the Paul trap
is expanded in terms of eigenstates of the LR invariant~\cite{Even}:
\begin{equation}
\Psi_{1x}(t)=\sum_{n=0,2,4,...}C_{1,n} \psi_{1,n} (x,t)
\end{equation}
($\Psi_{1y}$, $\Psi_{2x}$, and $\Psi_{2y}$ can be expanded
in a similar manner.)
When the electrons leave the Paul trap,
using Eq.~(\ref{BP}) or (\ref{qp:cp}), we have
\begin{eqnarray}
\Psi_{1x}(t+T)
= \sum_{n=0}^{\infty} C_{1,2n}\psi_{1,2n} (x,t)
e^{-i(2n+\frac{1}{2})\theta_{1x}(T)}.
\end{eqnarray}
Further, in the phase of each eigenstate,
the periodicity of the classical solution means that
$2n \theta_{1x}(T)$ is a multiple of $2\pi$.
Therefore, we can write
\begin{eqnarray}
\Psi_{1x}(t+T)&=&
e^{-\frac{i}{2} \theta_{1x}(T)}
\sum_{n=0}^{\infty} C_{1,2n}\psi_{1,2n} (x,t) \\
&=& e^{-\frac{i}{2} \theta_{1x}(T)}
\Psi_{1x}(t).
\end{eqnarray}
In the same way, we have
\begin{eqnarray}
\Psi_{1y}(t+T)
= e^{-\frac{i}{2} \theta_{1y}(T)}
\Psi_{1y}(t).
\end{eqnarray}
Then, we have the total wave function which travels path~1:
\begin{eqnarray}
\Psi_{1}(t+T)
= e^{-\frac{i}{2} [\theta_{1x}(T)+\theta_{1y}(T)]}e^{-i\theta_z}
\Psi_{1}(t),
\end{eqnarray}
for path 2, we have
\begin{eqnarray}
\Psi_{2}(t+T)
= e^{-\frac{i}{2} [\theta_{2x}(T)+\theta_{2y}(T)]}e^{-i\theta_z}
\Psi_{2}(t).
\end{eqnarray}
Therefore, the phase difference between two paths is
\begin{eqnarray}
\Theta_2-\Theta_1 &=&
\frac{1}{2}\{[\theta_{2x}(T)+\theta_{2y}(T)]
-[\theta_{1x}(T)+\theta_{1y}(T)]\}\\
&=& \theta_2(T) - \theta_1(T) .
\label{ph-dif}
\end{eqnarray}
Here, we have omitted the indices $x$ and $y$
since the phase changes for a period
are equal in the $x$- and the $y$-directions.

\subsection{Expected Results}

In this section, we present a typical experimental scheme.
In region III, we have the detector F, and
we have a destructive interference when
\begin{equation}
|\theta_2-\theta_1|=\pi.
\end{equation}
This destructive interference of the two wave functions, via
path 1 and 2, can be obtained by
controlling the applied voltage or the velocity of the electron beams.
By noting the fact that
when Eq.~(\ref{pe-cond}) holds,
the phase change over the minimal period is
\begin{equation}
\theta(\tau'=N'\tau/\epsilon)=l \pi ,
\label{theta}
\end{equation}
we have two methods to obtain destructive interference.
Firstly, we can control the applied voltages $\Phi_1$ and $\Phi_2$
so that $l=1$ in Eq.~(\ref{pe-cond}) and
$N'_1=2 N'_2$(where $N'_1$ and $N'_2$ are the values of $N'$
for path~1 and path 2, respectively).
Further, we can control the velocity of the electron beam so that
\begin{equation}
v = \frac{D}{N'_1 \tau}~(T=N'_1 \tau).
\label{vel}
\end{equation}
Then we have $\theta_1=\pi$ and $\theta_2=2\pi$.
Secondly, we can control the voltage values so that
$l=1$ and $l=2$ with the same $N'$.
From Eq.~(\ref{theta}), it is clear that two wave functions interfere
destructively.
In Table~1, we present the numerical values of $\omega_1$ and $\omega_2$
for $N'=4,8$ with $l=1$ and for $N'=3$ with $l=1,2$.

\begin{table}[bht]
Table 1.
Numerical values of $\omega_1$ and $\omega_2$
for $N\tau$-periodic CISs $(N=2,3,4)$
\begin{tabular}{cccccc}
\setlength{\tabcolsep}{5mm}
Fig.~\ref{fig4}
	 &$l$& $N'$& $\omega_1\tau$ = $\omega_2\tau$ & $\theta(\tau')$
\\[0.5ex] \hline
a& 1  & 4              & $3.14159$         & $\pi/2$ \\
b& 1  & 8              & $2.30517$         & $\pi/2$ \\
c& 1  & 3              & $2.63690$         & $\pi/2$ \\
d& 2  & 3              & $3.48328$         & $\pi $  \\
\end{tabular}
\end{table}
The graphs of $g_{-}(t)$ in the $x$-direction for all the cases
in Table 1 are shown in Fig.~4.
The parameters are fixed
as in Eq.~(\ref{const}), and hence the region of $g_- (t) = 1/m$
(the figures are depicted in $m=1$ units)
reflects that $I(t)=H(t)$, as discussed in Ref.~\cite{Ji95:Exac}.
By shifting these figures by a half period, $\tau/2$,
we can also get $g_-(t)$ in the $y$-direction.
As expected, they are $N\tau$-periodic ($N=2,3,4$),
as are their corresponding
wave functions, Eq.~(\ref{WF}).
The original and the shifted figures also reveal that the probability density
function $|\Psi (x,y,t)|^2$ spreads in the $x$-direction
and the $y$-direction alternately.
\begin{figure}[htb]
\centerline{\epsfig{figure=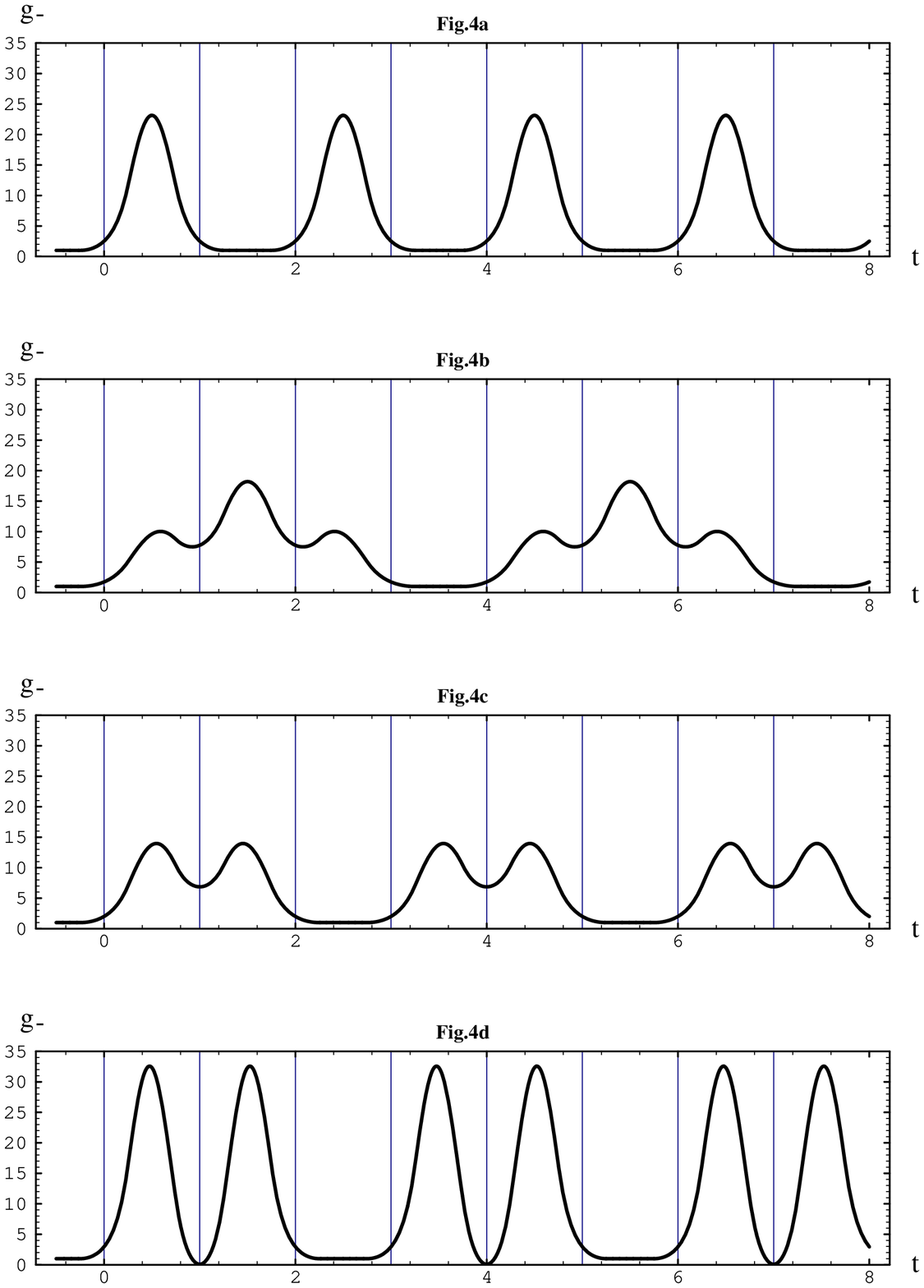,width=120mm,angle=0}}
\caption{The shapes of $g_-(t)$ for Table 1.
The time is denoted in units of $\tau$.
}
\label{fig4}
\end{figure}

\section{Discussion}

Modification of the time-dependent electric potential
of the Paul trap from a sinusoidal waveform to a square waveform
gives a quantum solution with a simple mathematical form.
Therefore, we can verify the existence of
the $N\tau$-periodic CIS and
propose a method to detect the corresponding Berry's phase
by experiment.

We estimate the values of the parameters for a practical experiment.
To obtain an interference pattern successfully, we should have a Paul trap of
submillimeter size ($d \sim 10^{-3}~{\rm m}$).
Considering a Paul trap whose length $D$ is of the order of $10^{-1}~{\rm m}$
 and an electron with a speed on the order of $10^6~{\rm m/s}$,
$\tau' \sim 10^{-7}~{\rm s}$,
we have $\omega_1 \sim 10^8~{\rm s^{-1}}$  from Table 1 and
$\Phi \sim 1~{\rm V}$ from Eq.~(\ref{freq12}).
For example, for $N'=4$, $l=1$ in Table~1,
$D \approx 6~{\rm cm}$,
$d \approx 1~{\rm mm}$, and $v \approx 5\times10^6~{\rm m/s}$,
we have
$\tau \approx 6\times 10^{-9}~{\rm s}$,
$\omega_{(1,2)} \approx 5\times 10^8~{\rm s^{-1}}$,
and $|\Phi_{(1,2)}| \approx 1~{\rm V}$.
These values seem practical for an experimental arrangement.

There have been many applications and tests
of the Berry's phase using an optical fiber~\cite{TomitaC86},
nuclear magnetic resonance (NMR)~\cite{SuterMP88},
etc.~\cite{ShapereW89}.
Nonetheless, there are no experiments about the Berry's phase
caused by the quantum motions in phase space.
(Note that the optical phase effect deals with the phenomena of
classical electromagnetism, and the NMR experiment investigates
the interaction between the spin and the external magnetic fields.)
Our proposal will be a new experiment to detect the Berry's phase
caused by a pure dynamics in phase space,
and we expect that it will play a significant role in understanding
the quantum motions of a time-dependent system in phase space.

\section{Acknowledgments}
This work was supported by the Center for Theoretical Physics at Seoul National University
and by the Basic Science Research Institute Program, Ministry of Education,
Project No. BSRI-96-2418.

\newpage
\begin{center}
{\large
Figure captions
}
\end{center}

\begin{itemize}
\item Fig. 1. The time-dependent potential of a square waveform.

\item Fig. 2. The stability-instability diagram.
The vertical strips stand for the stable regions in the $x$-motion,
the horizontal strips for the $y$-motion.

\item Fig. 3. The schematic diagram of
the experimental arrangement.

\item Fig. 4. The shapes of $g_-(t)$ for Table 1.
The time is denoted in units of $\tau$.

\end{itemize}

\begin{references}
\bibitem{Paul90} W. Paul, Rev. Mod. Phys. {\bf 62}, 531 (1990).
\bibitem{Brown91} L. S. Brown, Phys. Rev. Lett. {\bf 66}, 527 (1991).
\bibitem{FengW95} M. Feng and K. Wang, Phys. Lett. A {\bf 197}, 135 (1995).
\bibitem{Ji95:Heis} J. Y. Ji, J. K. Kim, and S. P. Kim,
Phys. Rev. A {\bf51}, 4268 (1995).
\bibitem{LR69} H. R. Lewis, Jr., and W. B. Riesenfeld, J. Math. Phys.
{\bf 10}, 1458 (1969).
\bibitem{Ji95:Exac} J. Y. Ji,  J. K. Kim,  S. P. Kim, and K.  S. Soh,
Phys.  Rev. A  {\bf52}, 3352 (1995).
\bibitem{Berry84} M. V. Berry, Proc. R. Soc. Lond. A {\bf 392}, 45 (1984).
\bibitem{AharonovA87} Y. Aharonov and J. Anandan,
Phys. Rev. Lett. {\bf 58}, 1593 (1987).
\bibitem{Ji95:Ntau} J. Y. Ji and J. K. Kim,
Phys. Lett. A {\bf 208}, 25 (1995).
\bibitem{Ji96:Temp} J. Y. Ji and J. K. Kim, Phys. Rev. A {\bf 53}, 703 (1996).
\bibitem{MagnusW66} W. Magnus and S. Winkler,
{\it Hill's Equation } (Dover, New York, 1966).
\bibitem{AharonovB59} Y. Aharonov and D. Bohm,
Phys. Rev. {\bf 115}, 485 (1959).
\bibitem{Even}
The plane wave can be considered as the limiting case
of a particle in a box and the expansion coefficient
$c_n=\lim_{L \rightarrow \infty} \int_{-\frac{L}{2}}^{\frac{L}{2}}
\frac{1}{\sqrt{L}} \psi_n (q,t)dq~(q = x,y)$
vanishes for odd $n$.
\bibitem{TomitaC86} A. Tomita and R. Chiao, Phys. Rev. Lett. {\bf 57},
937 (1986).
\bibitem{SuterMP88} D. Suter, K. T. Mueller, and A. Pines,
Phys. Rev. Lett. {\bf 60}, 1218 (1988).
\bibitem{ShapereW89} For more details, see {\it Geometric Phases in Physics},
editted by A. Shapere and F. Wilczek (World Scientific, Singapore, 1989).

\end{references}
\end{document}